# Resolution Dependence of Disruptive Collisions between Planetesimals in the Gravity Regime

Hidenori Genda[1*], Tomoaki Fujita[2], Hiroshi Kobayashi[3], Hidekazu Tanaka[4], Yutaka Abe[2]

[1]Earth-Life Science Institute, Tokyo Institute of Technology,
Ookayama, Meguro-ku, Tokyo, 152-8550, Japan

[2]Department of Earth and Planetary Science, The University of Tokyo,
Hongo, Bunkyo-ku, Tokyo, 113-0033, Japan

[3]Department of Physics, Graduate School of Science, Nagoya University,
Furo-cho, Chikusa-ku, Nagoya, 464-8602, Japan

[4]Institute of Low Temperature Science, Hokkaido University,
Sapporo 060-0819, Japan

*Corresponding author: Hidenori Genda

Mailing address: Earth-Life Science Institute, Tokyo Institute of Technology,
2-12-1-IE-10 Ookayama, Meguro-ku, Tokyo, 152-8551, Japan

Email: genda@elsi.jp, Phone: +81-3-5734-2887

# Abstract

Collisions are a fundamental process in planet formation. If colliding objects simply merge, a planetary object can grow. However, if the collision is disruptive, planetary growth is prevented. Therefore, the impact conditions under which collisions are destructive are important in understanding planet formation. So far, the critical specific impact energy for a disruptive collision $Q_\mathrm{D}^*$ has been investigated for various types of collisions between objects ranging in scale from centimeters to thousands of kilometers. Although the values of $Q_\mathrm{D}^*$ have been calculated numerically while taking into consideration various physical properties such as self-gravity, material strength, and porosity, the dependence of $Q_\mathrm{D}^*$ on numerical resolution has not been sufficiently investigated. In this paper, using the smoothed particle hydrodynamics (SPH) method, we performed numerical simulations of collisions between planetesimals at various numerical resolutions (from $2 \times 10^4$ to $5 \times 10^6$ SPH particles) and investigated the resulting variation in $Q_\mathrm{D}^*$. The value of $Q_\mathrm{D}^*$ is shown to decrease as the number of SPH particles increases, and the difference between the $Q_\mathrm{D}^*$ values for the lowest and highest investigated resolutions is approximately a factor of two. Although the results for $5 \times 10^6$ SPH particles do not fully converge, higher-resolution simulations near the impact site show that the value of $Q_\mathrm{D}^*$ for the case with $5 \times 10^6$ SPH particles is close to the expected converged value. Although $Q_\mathrm{D}^*$ depends on impact parameters and material parameters, our results indicate that at least $5 \times 10^6$ SPH particles are required for numerical simulations in disruptive collisions to obtain the value of $Q_\mathrm{D}^*$ within 20% error.

# 1. Introduction

It is generally accepted that planets grow in protoplanetary disks composed of dust and gas (e.g., Hayashi et al., 1985; Ida and Lin, 2004). The process of fine dust growing into a planet can be divided into several stages. The first stage is characterized by the accumulation of dust and the formation of planetesimals, which are typically 1–100 km in size (Safronov, 1969; Wetherill, 1980; Goldreich and Ward, 1973; Weidenschilling, 1980, 1984; Wada et al., 2007, 2008, 2009; Kataoka et al., 2013). In the next stage, planetesimals collide with each other and grow (Greenberg et al., 1978; Wetherill and Stewart, 1989; Kokubo and Ida, 1996). In the terrestrial planet region (inside the ice line), several tens of Mars-sized rocky protoplanets are formed. In the gas giant planet region, because there are so many icy planetesimals, very large icy protoplanets whose masses are several times that of Earth are formed. In the last stage of gas giant planet formation, such large protoplanets begin to rapidly capture the surrounding nebular gas. Ultimately, gas giant planets such as Jupiter or Saturn are formed. In the final stage of terrestrial planet formation, Mars-sized protoplanets frequently collide with each other, ultimately forming Earth-sized terrestrial planets (Chambers and Wetherill, 1998; Agnor et al., 1999; Kokubo and Genda, 2010; Genda et al., 2012).

Many collisions constantly take place during planet formation. If colliding bodies merge, collisions promote planet growth. However, collisions do not always promote growth. For example, in the stage of planetesimal formation, collisions between dust aggregates accelerated by turbulence in a protoplanetary disk can be so destructive that the dust aggregates break into fragments instead of growing (Weidenschilling, 1984; Wada et al., 2008). Additionally, the stage of protoplanet formation involves a similar problem. Once protoplanets become massive, their stirring increases the random velocity of surrounding planetesimals, and collisions between planetesimals become more destructive. As the fragments resulting from the destructive collisions between planetesimals are removed by rapid radial drift due to gas drag in the protoplanetary disk, the depletion of bodies accreting onto protoplanets stalls protoplanet growth (Inaba et al., 2003; Kenyon and Bromley, 2008; Kobayashi et al., 2010, 2011). Conversely, the radial drift of fragments resulting from destructive collisions accelerates protoplanet growth at a pressure maximum in the protoplanetary disk (Kobayashi et al., 2012). Therefore, the conditions of collisional destruction for planetesimals is very important in understanding planet formation.

Impact energy is presumed to greatly influence collision outcomes and is useful to estimate how destructive a collision is. If the larger colliding body, target, is much larger than the smaller one, impactor, the specific impact energy $Q$ is given by $Q = E_{imp}/M_{tar}$, where $E_{imp}$ and $M_{tar}$ are the impact energy and the target mass, respectively. The impact energy $E_{imp}$ is given by $E_{imp} = 0.5\ m_{imp}\ v_{imp}^2$, where $v_{imp}$ and $m_{imp}$ are the impact velocity and the impactor mass ($M_{tar} > m_{imp}$), respectively. The critical specific impact energy $Q_D^*$, which is the specific impact energy required to disperse the target in two or more bodies with the largest body having exactly half the mass of the original target (i.e., $M_{tar}/2$) after the collision, is often used to characterize disruptive collisions. The value of $Q_D^*$ for planetesimals determines the timescales of the collisional evolution of planetesimal swarms and debris disks (e.g., Wyatt et al., 2007; Kobayashi and Tanaka, 2010), which are related to planet formation (Kobayashi and Löhne, 2014).

When $m_{imp}$ is not much smaller than $M_{tar}$, $Q_{RD}^*$ should be used instead of $Q_D^*$ (e.g., Leinhardt and Stewart 2012), because $Q_{RD}^*$ includes the size effect of the impactor. However, in our all numerical simulations, $m_{imp}$ is less than 2% of $M_{tar}$ (see Section 2.2), which means that $Q_{RD}^*$ is almost identical to $Q_D^*$.

$Q_D^*$ has been investigated using various approaches: laboratory experiments (e.g., Housen and Holsapple, 1999; Holsapple et al., 2002; Nakamura et al. 2009), analytical or scaling methods (e.g., Housen and Holsapple, 1990; Mizutani et al. 1990), asteroid belt observations (Durda et al., 1998), and numerical calculations (Love and Ahrens, 1996; Melosh and Ryan, 1997; Benz and Asphaug, 1999; Leinhardt and Stewart, 2009, 2012; Jutzi et al. 2010, Jutzi 2015). For large-scale collisions, such as collisions between planetesimals or protoplanets, numerical calculations have been powerful tools to investigate collision phenomena because direct experimental measurements of such collisions are difficult in the laboratory.

In the gravity regime (target radius $R_{tar} > \sim 1$ km), $Q_D^*$ increases with $R_{tar}$. The value of $Q_D^*$ in the gravity regime has been calculated by several numerical methods so far (Figure 1). One frequently used method is the smoothed particle hydrodynamics (SPH) method (Love and Ahrens, 1996; Benz and Asphaug, 1999; Jutzi et al., 2010; Jutzi, 2015), which is a Lagrangian method used to solve fluid motion (e.g., Monaghan, 1992; Springel, 2010). The other methods are the two-dimensional Lagrangian hydrocode (Melosh and Ryan, 1997), the hybrid code of the Eulerian hydrocode and the *N*-body code (Leinhardt and Stewart, 2009, 2012), and direct *N*-body code (e.g., Leinhardt and Richardson 2002; Leinhardt et al. 2012). Among these methods, there are variations in the value of $Q_D^*$ for constant $R_{tar}$ by up to a factor of 10, as shown in Figure 1. Leinhardt and Stewart (2012) could explain the spread of $Q_D^*$ among the previous data to some extend by applying the scaling laws of the impact conditions (mass ratio, impact velocity, and impact angle) and material properties (strength and density). However, among the SPH methods (Love and Ahrens, 1996; Benz and Asphaug, 1999; Jutzi et al., 2010), there are also variations in the value of $Q_D^*$, which seems to be caused by different physical properties and material parameters among these codes. The code developed by Love and Ahrens (1999) includes self-gravity but not material strength, whereas that by Benz and Asphaug (1999) includes material strength but not self-gravity. That by Jutzi et al. (2010) includes both self-gravity and material strength, but their parameters for material strength are different from those used in Benz and Asphaug (1999).

The number of SPH particles used for collisions also differs among these studies. Love and Ahrens (1999) used $2 \times 10^3$ SPH particles, Benz and Asphaug (1999) used $5 \times 10^4$ SPH particles, and Jutzi et al. (2010) used $2 \times 10^5$ SPH particles. However, the dependence of $Q_D^*$ on the numerical resolution has not yet been sufficiently investigated. Since the included physics and parameters differ among these codes, the resolution dependence of $Q_D^*$ cannot be directly discussed in their results. For example, when $1 \times 10^5$ SPH particles are used for a collisional simulation with a typical target-to-impactor mass ratio $M_{tar}/m_{imp} \sim 100$, the impactor consists of only ~1000 SPH particles. Then the diameter of the impactor is resolved with only ~10 SPH particles. It should be clarified how large error arises in $Q_D^*$ from such a marginal resolution.

In this paper, we use SPH to perform collisional simulations for various resolutions with the same impact parameters and the same code, and investigate the resolution dependence of $Q_D^*$ for collisions in the gravity regime. To focus on the resolution dependence, we include self-gravity but not material strength in our code.

In Section 2, the methods for the numerical code and analysis of collision outcomes are introduced. Section 3 presents the numerical simulation results and investigates the dependence of $Q_D^*$ on the numerical resolution. We discuss this dependence in Section 4.

# 2. Methods

## 2.1. Numerical Code for Collisions

To perform impact simulations of planetesimals, we use the SPH method (e.g., Monaghan, 1992), which is a flexible Lagrangian method of solving hydrodynamic equations and has been widely used for impact simulations in planetary science. The SPH method can easily process large deformations and shock waves. Our numerical code is based on the works of Genda et al. (2012) and Sekine and Genda (2012), but we use a modified version. Here, we briefly describe the essential points of the code.

For the previous version of our SPH code, the mutual gravity between all SPH particles was directly computed using a special-purpose computer for gravitational $N$-body systems named GRAPE-6A (Fukushige et al., 2005). The computational cost is proportional to $N^2$, where $N$ is the number of particles. In the new version, the mutual gravity is calculated using the standard Barnes–Hut tree method (Barnes and Hut, 1986; Hernquist and Katz, 1989) on a multicore central processing unit (CPU). The computational cost is proportional to $N\log N$ to allow us to deal with the large number of SPH particles. Additionally, we apply modified terms in the equations of motion and energy proposed by Price and Monaghan (2007) to more effectively conserve the energy. In our all simulations, the error of the total energy is within 0.1% during impact simulation.

The Tillotson equation of state (Tillotson, 1962) is used in our SPH code and has been widely applied in other previous studies including planet- and planetesimal-sized collisional simulations (e.g., Benz and Asphaug, 1999; Canup and Asphaug, 2001; Jutzi et al., 2010; Genda et al., 2012; Hosono et al., 2013; Citron et al., 2015). The Tillotson equation of state contains ten material parameters, and the pressure is expressed as a function of the density and the specific internal energy, which is convenient for treating fluid dynamics. We used the parameter sets of basalt referenced in Benz and Asphaug (1999).

In our code, we use a Von Neumann–Richtmyer-type viscosity with parameters of $\alpha = 1.5$ and $\beta = 3.0$ for standard cases, as described in Monaghan (1992). We discuss the effect of the values of $\alpha$ and $\beta$ in Section 3.3.

## 2.2. Initial Conditions for Collisions

As shown in Figure 2, we simulate impacts between two planetesimal-sized spheres. In this paper, we consider two types of targets with $R_{tar}$ = 10 and 100 km. We also consider the cases for two impact angles, $\theta = 0°$ and 45°, where $\theta = 0°$ corresponds to a head-on impact and $\theta = 90°$ corresponds to a grazing impact. In all of our simulations, the impact velocity is fixed at 3 km/s, which is the same as that used in Benz and Asphaug (1999). To obtain the value of $Q_D^*$, we need to perform several impact simulations with different impact energies $Q$. We consider five impactor sizes in the range of approximately 0.001–0.1 $M_{tar}$, depending on the target size and impact angle. We calculate until 300–600 s after impacts.

Both the target and impactor are made of basalt. Particles are placed on grid points of a mesh with a density of 2700 kg/m$^3$ within a sphere of a fixed size. The number of SPH particles $n_{tar}$, which corresponds to the numerical resolution, that comprise the target is $5 \times 10^4$, $1 \times 10^5$, $5 \times 10^5$, and $5 \times 10^6$ SPH particles. All SPH particles in a target and an impactor have the same mass. For example, the number of SPH particles in an impactor for a target with $5 \times 10^4$ particles is 100–1000 particles depending on the impactor mass. The case of $5 \times 10^4$ target particles, which is the minimum resolution in our simulations, is the same as that used in Benz and Asphaug (1999). It takes several days to perform one impact simulation with the highest resolution ($5 \times 10^6$ particles) using a computer with 64 cores (4 CPU, AMD Opteron™ Processor 6276, 2.3 GHz).

## 2.3. Analysis of the Mass of the Largest Body

To obtain the value of $Q_D^*$, the mass of the largest body $M_{lrg}$ should be calculated from the collision outcome data. Like the procedure in Benz and Asphaug (1999), we calculate $M_{lrg}$ using the following three-step procedure. First, we use a friends-of-friends (FOF) algorithm to identify clumps of SPH particles. If the distance between two particles is less than a certain threshold value $l_{FOF}$, these particles are defined as belonging to the same clump. The distances for all pairs of particles are evaluated. The value of $l_{FOF}$ is set to be slightly larger than the typical distance between two nearest SPH particles under initial conditions. In this way, we roughly identify clumps and call them FOF groups. There are also numerous SPH particles that do not belong to any FOF groups. We call these particles isolated particles.

Next, we determine if the particles in an FOF group are gravitationally bound. If the kinetic energy of the $j$-th particle in the FOF group (½$v_j^2$, where $v_j$ is the velocity of the $j$-th particle) is larger than the gravitational potential energy of the group ($GM_{FOF}/r_j$, where $G$ is the gravitational constant, $M_{FOF}$ is the mass of the FOF group, and $r_j$ is the distance between the $j$-th particle and the center of mass of the FOF group), the $j$-th particle is removed from the FOF group. We also determine if the isolated particles are gravitationally bound to each FOF group. This procedure is iteratively performed until the particle numbers of the FOF groups converge. We call the converged FOF groups singly gravitationally bound (SGB) groups.

Finally, we iteratively determine if each SGB group is gravitationally bound. If two SGB groups are gravitationally bound to each other, we regard them as a single group called a finally gravitationally bound (FGB) group. We define the mass of the largest FGB group as the mass of the largest body $M_{lrg}$. Using this procedure, the value of $M_{lrg}$ quickly converges after the passage of shock and rarefaction waves in the target body. It enables us to obtain $M_{lrg}$ with a relatively short term run with $t \sim 300$ s (see Section 3.1 for details).

## 3. Collision Outcomes
### 3.1. Typical Results for Collisions

Figure 3 shows snapshots of a cross section of one typical simulated head-on impact. The radii of the target and impactor are 100 and 15 km, respectively. The impactor collides with the target at a velocity of 3 km/s. The target and impactor consist of $5 \times 10^6$ and $2 \times 10^4$ particles, respectively. The color contour in Figure 3 represents the specific kinetic energy. The snapshots show how the shock wave propagates from the impact site to the rear of the target sphere and how the ejecta are scattered. First contact occurs at $t = 0$ s, and the isobaric core is formed at approximately $t = 8$ s. After that, the initial shock wave arrives at the rear of the target at $t = 56$ s. Ejection continues until $t \approx 300$ s.

Figure 4 shows the $M_{lrg}$ evolution for the impact simulation shown in Figure 3. As shown in this figure, $M_{lrg}/M_{tar}$ almost converges to 0.76 within a short time (~150 s) after the impact. This is because the internal and kinetic energies of each particle do not change considerably, because of no strong interaction between particles after the shock and rarefaction waves propagate. The propagation is completed in approximately 100 s. Although the re-accretion of ejected SPH particles onto the largest body occurs on a dynamical timescale ($\sqrt{3\pi/16G\rho} \approx 2000$ s, where $\rho$ is the typical density of the target), we can confirm that $M_{lrg}$ quickly converges in a short time ($\approx 150$ s) as a result of the detailed procedure of determining $M_{lrg}$ in Section 2.3. We also confirmed this quick convergence by carrying out some impact simulations with $5 \times 10^5$ particles for 5000 s.

Figure 5 shows snapshots of one oblique impact ($\theta = 45°$). The radii of the target and impactor are 100 and 28 km, respectively. The number of particles composing the target is $n_{tar} = 5 \times 10^6$. The mass of the largest body after the impact is $M_{lrg}/M_{tar} = 0.6$. A part of the impactor does not strongly interact with the target but moves away from it.

## 3.2. Determination of $Q_D^*$

One series of calculations (one SPH impact simulation and data analysis) gives the mass of the largest body under one set of impact conditions (target size, impactor size, impact velocity, and impact angle). Figure 6 shows the collision outcomes of impact simulations for various impactor masses where $R_{tar}$, $v_{imp}$, and $\theta$ are fixed at 100 km, 3 km/s, and 0°, respectively. The second data point from the left in Figure 6 represents the impact simulation shown in Figures 3 and 4. Changes in the impactor mass correspond to changes in the impact energy $E_{imp}$.

As expected, collisions with higher impact energies are more disruptive, i.e., lower $M_{lrg}$. Under this impact condition, the critical specific impact energy for disruptive collision $Q_D^*$ where the mass of the largest body is exactly half the mass of the original target can be obtained by linear interpolation, yielding $Q_D^* = 31$ kJ/kg. In the case of $n_{tar} = 5 \times 10^6$, the number of SPH particles in the impactor is $3.3 \times 10^4$ for $Q = 30$ kJ/kg.

## 3.3. Dependence of $Q_D^*$ on Numerical Resolution

Here, we investigate the dependence of $Q_D^*$ on the number of SPH particles used in head-on impact simulations. Figure 7 shows $Q_D^*$ for $n_{tar} = 5 \times 10^4$, $1 \times 10^5$, $5 \times 10^5$, and $5 \times 10^6$, where $n_{tar}$ is the number of particles in the target. The value of $Q_D^*$ decreases as the number of particles increases. $Q_D^*$ in the case of $n_{tar} = 5 \times 10^6$ is less than two-thirds of that in the case of $n_{tar} = 5 \times 10^4$. The numerical resolution of $5 \times 10^4$ particles is almost the same as that used in Benz and Asphaug (1999). It is clear that the impact simulation with $5 \times 10^4$ SPH particles is insufficient for determining $Q_D^*$.

The resolution dependence of $Q_D^*$ is well fit by

$$Q_D^* = a + bn_{\text{tar}}^{-1/3}, \qquad (1)$$

where $a$ and $b$ are fitting parameters. The value of $a$ in Eq. (1) corresponds to the converged $Q_D^*$ in the limit of $n_{\text{tar}} \to \infty$. Using four values of $Q_D^*$ obtained by our numerical simulations, the fitting parameters are determined to be $a$ = 26.8 kJ/kg and $b$ = 782 kJ/kg. The value of $Q_D^*$ (= 31 kJ/kg) for $n_{\text{tar}} = 5 \times 10^6$ is only 16% larger than the expected converged value. Although the standard SPH code used here is second order accurate in space, $Q_D^*$ converges with $n_{\text{tar}}^{-1/3}$ instead of $n_{\text{tar}}^{-2/3}$. Since the diameter of the target is resolved with ~ $n_{\text{tar}}^{1/3}$ SPH particles, $n_{\text{tar}}^{-1/3}$ corresponds to the initial distance among the nearest SPH particles (~ a smoothing length $h$). The conversion with $n_{\text{tar}}^{-2/3}$ (~ $h^2$) occurs if the physical values such as density, pressure and internal energy are differentiable in space. However, our impact simulations involve shock waves, and the physical values at shock front are not differentiable: The shock front can be resolved by several SPH particles (i.e., ~ $h$) by using artificial viscosity. Therefore, it is reasonable that the results converge with $h \sim n_{\text{tar}}^{-1/3}$.

If we choose $n_{\text{tar}} = 2 \times 10^4$, $Q_D^*$ is approximately twice as large as the converged value. To further check the resolution convergence, we should perform much higher-resolution simulations. However, these would cost a large amount of CPU time. In this paper, instead of performing such higher-resolution simulations, we investigate the dependence of the energy distribution on the numerical resolution in the following section. There, we find that resolution dependence occurs in a short time after impact. In Section 4.2, we perform impact simulations very close to an impact site with a much higher resolution.

In addition to head-on impact simulations, we performed oblique impact simulations ($\theta$ = 45°) with different values of $Q$. We found that $Q_D^*$ for the oblique impact with $n_{\text{tar}} = 5 \times 10^6$ is 120 kJ/kg for $R_{\text{tar}}$ = 100 km and 6.0 kJ/kg for $R_{\text{tar}}$ = 10 km. These values are expected to be close to the converged ones, as in the head-on impact cases. The obtained values of $Q_D^*$ for the head-on and oblique collisions with $n_{\text{tar}} = 5 \times 10^6$ are plotted in Figure 1 with the previously reported values of $Q_D^*$.

We also investigated the effect of artificial viscosity on $Q_D^*$. Figure 8 shows the dependence of $Q_D^*$ on the number of SPH particles for different values of the coefficient for a Von Neumann-Richtmyer-type viscosity. The value of $Q_D^*$ depends on the artificial viscosity. Higher viscosity ($\alpha$ = 2.0 and $\beta$ = 4.0) tends to be higher $Q_D^*$, which makes sense because more impact energy is transferred to thermal (internal) energy.

However, $Q_\mathrm{D}^*$ seems to converge to almost same values independent of artificial viscosity at high resolution.

## 4. Discussion of Resolution Dependence
### 4.1. Energy Transfer During an Impact

Here, we focus on the evolution of the total kinetic energy $K$ and the total internal energy $U$ during the collision, which are respectively defined by

$$K = \sum_j \frac{1}{2} m^j (v^j)^2,$$

$$U = \sum_j m^j u^j,$$

where $m^j$, $v^j$, and $u^j$ are the mass, the velocity, and the specific internal energy of the $j$-th SPH particle, respectively. Here, we define $E = K + U$, and $E$ does not include the potential energy. Figure 9 shows the temporal evolution of $K$ and $U$ for the impact simulation shown in Figure 3. As shown in Figure 9, the total kinetic energy decreases, and the total internal energy increases steeply for a short time just after the impact. The transfer from kinetic to internal energy is mainly caused by the impact shock wave. During the passage of the shock and rarefaction waves, the impact energy (i.e., kinetic energy) is distributed among the internal energy, the kinetic energy of the ejecta, and the gravitational potential. The evolution of the total kinetic energy clearly depends on the resolution, which means that the efficiency of the energy transfer depends on the resolution. Higher total kinetic energy remains after several hundred seconds for the case of higher-resolution simulations, and lower total kinetic energy remains for the case of lower-resolution simulations. This dependence is consistent with the resultant $Q_\mathrm{D}^*$ shown in Figure 7 because more particles can escape if they have more kinetic energy.

As shown in Figure 9(a) and (b), most of the energy transfer from kinetic to internal energy takes place within ~50 s. The efficiency of this energy transfer within ~50 s is higher for lower-resolution simulations. The remarkable difference between the efficiencies of the different resolution cases appears within the first 15 s (see Figure 9(c)). Therefore, to confirm the convergence of the energy transfer from kinetic to internal energy, we do not need to calculate the whole target over a long period. Instead, we perform impact simulations only near the impact site for the shorter 15 s period, as described in the next section.

## 4.2. Impact Simulation Near an Impact Site

A remarkable difference between the efficiencies of the energy transfer from kinetic to internal energy for different numerical resolutions appears within the first 15 s. Therefore, precisely solving the energy transfer near the impact site is the key to confirming numerical convergence. To check the dependence of the energy transfer for higher-resolution simulations, we perform impact simulations only near the impact site and over a short time. Thus, the calculation cost is reduced because the region required for calculation is only part of the sphere rather than the whole sphere. Moreover, the calculation time is shorter than that for a sphere-to-sphere collision (~300 s).

For the target body, we prepare a bowl with a curvature. Figure 10 shows snapshots of a cross section of a typical sphere-to-bowl impact simulation. The impact parameters are $R_{\mathrm{imp}}$ = 15.8 km and $v_{\mathrm{imp}}$ = 3.0 km/s. The curvature of the target body is set to be 100 km. The color contour represents the specific kinetic and specific internal energies in the left and right groups of four snapshots, respectively.

The volume of the bowl-shaped target body is approximately 10% of that of the perfectly spherical target, and the calculation time is only 15 s. These two benefits allow us to perform impact simulations with a much higher resolution within a practical time. We used approximately $7.5 \times 10^4$–$1.5 \times 10^7$ particles for the bowl-shaped target, which corresponds to approximately $7.5 \times 10^5$–$1.5 \times 10^8$ particles for the spherical target.

Figure 11 shows the evolution of the total kinetic energy for the sphere-to-bowl impact simulations during the first 15 s and the value of the total kinetic energy at $t = 15$ s as a function of the number of the sphere-equivalent particles. More total kinetic energy remains at higher-resolution impacts, which is similar to the sphere-to-sphere impact results shown in Figure 9. As with the fitting of $Q_{\mathrm{D}}^*$ in Equation (1), the resolution dependence of the final value of the total kinetic energy at 15 s ($[K/E]_{t=15\mathrm{s}}$) is fit by

$$[K/E]_{t=15\mathrm{s}} = c + d n_{\mathrm{tar}}^{-1/3}. \qquad (2)$$

The fitting parameters are determined to be $c$ = 0.79 and $d$ = −11.7. The evolution of $K/E$ for $7.5 \times 10^7$ sphere-equivalent particles appears to be almost identical to that for $1.5 \times 10^8$ sphere-equivalent particles, as shown in Figure 11(a). However, Figure 11(b) indicates that more than $1.5 \times 10^8$ sphere-equivalent particles may be required for convergence. Figure 11(b) also shows that this fitting function works well even at higher resolutions with $n_{\text{tar}} > 5.0 \times 10^6$ particles (or $n_{\text{tar}}^{-1/3} < 0.006$). In our simulations, we considered the collisions with $v_{\text{imp}} = 3$ km/s. For protoplanet formation, we can apply this discussion, because typical collision velocities among planetesimals excited by surrounding Moon-sized or Mars-sized protoplanets are ~ 3 km/s. However, we should be careful for the case of higher-velocity impact. It is expected that the number of SPH particles needed for convergence would be depend on the impact velocity, because more particles would be needed to precisely solve stronger shock wave and to resolve a smaller impactor with the sufficient number of SPH particles.

## 5. Summary

In this paper, by using the SPH method with self-gravity and without material strength, we performed numerical simulations for head-on and oblique ($\theta = 45°$) collisions between a basaltic impactor and target ($R_{\text{tar}} = 10$ and $100$ km) with $v_{\text{imp}} = 3$ km/s. Varying the number of SPH particles from $2 \times 10^4$ to $5 \times 10^6$, we investigated the dependence of $Q_D^*$ on the numerical resolution. We found that the value of $Q_D^*$ decreases as the number of SPH particles increases, and the difference between the $Q_D^*$ values at the lowest and highest investigated resolutions is approximately a factor of two. This difference is caused by the different efficiencies of the energy transfer from kinetic to internal energy during the propagation of the shock and rarefaction waves through the impactor and target.

Although the value of $Q_D^*$ for $5 \times 10^6$ SPH particles does not fully converge, the fitting curve of $Q_D^*$ shown in Figure 7 indicates that it is close to the expected converged value. Local simulations performed near the impact site with higher resolutions also support this idea. Although $Q_D^*$ depends on impact parameters and material parameters, our results indicate that at least $5 \times 10^6$ SPH particles are required for numerical simulations in disruptive collisions in the gravity regime to obtain the value of $Q_D^*$ within 20% error. Previous studies used numbers of SPH particles ranging from $2 \times 10^3$–$2 \times 10^5$ (Love and Ahrens 1996; Benz and Asphaug 1999; Jutzi et al. 2010; Jutzi 2015), which overestimated the values for $Q_D^*$ by the factor of 2. In context, $5 \times 10^6$ SPH particles is not a huge number, because the target body is composed of

~$200^3$ particles, which means that only 200 particles are allocated in one dimension. To roughly capture the propagation of the shock and rarefaction waves, at least 10 particles are required for one-dimensional calculations. Thus, more than 100 particles are needed to precisely calculate the energy transfer during the propagation of the shock and rarefaction waves.

Our work shows that SPH collision simulations do converge with increasingly high resolution, and moderate resolution (a few $10^5$ SPH particles) are within ~ 50% of the convergence limit. Therefore, we do not always need hyper-resolution simulations (a few $10^6$ SPH particles) for large parameter space studies of impact simulations. However, because the error in $Q_D^*$ would depend on the impact conditions such as the impact velocity and material strength, we sometimes have to carry out hyper-resolution simulations in order to check convergence. In our impact simulations, we used the equal-mass SPH particles. If different-mass SPH particles are used, for example, a lot of small mass SPH particles are assigned to the impactor and contact region in the target, we can save the computational time to keep the accuracy of $Q_D^*$. The usage of the SPH with particle splitting (e.g., Kitsionas and Whitworth 2002) is also better way to deal with this problem.

The value of $Q_D^*$ has been applied in studies of the evolution of the asteroid belt (e.g., Bottke et al., 2005), debris disk formation (Wyatt et al., 2007), and planet formation (Kobayashi et al., 2010; Kobayashi et al., 2010, 2011; Kenyon and Bromley, 2012). For example, the value of $Q_D^*$ directly affects the mass of formed protoplanets, which grow through successive collisions of planetesimals in the protoplanetary disk. According to Kobayashi et al. (2010), the mass of formed protoplanets is proportional to $(Q_D^*)^{0.87}$. A factor of two difference in $Q_D^*$ directly affects the mass of protoplanets by a factor of 1.8. The mass of protoplanets is directly related to the formation of Mercury and Mars, as they are thought to be survivors of protoplanets (Kobayashi and Dauphas, 2013). Moreover, the mass of protoplanets is the most important factor in runaway gas accretion forming gas giant planets such as Jupiter, Saturn, and many extrasolar planets. Although the effects of impact parameters and material strength on $Q_D^*$ are quite large, high-resolution simulations for determining $Q_D^*$ are also important in understanding planet formation.

# Acknowledgments

We thank anonymous reviewers for valuable comments and suggestions on our manuscript. This work was supported by JSPS KAKENHI Grant Numbers 26287101 and 15K13562, and Research Grant 2015 of Kurita Water and Environment Foundation.

# References

Agnor, C.B., Canup, R.M., Levison, H.F., 1999. On the character and consequences of large impacts in the late stage of terrestrial planet formation. Icarus 142, 219–237.

Barnes, J., Hut, P., 1986. A hierarchical O (N log N) force-calculation algorithm. Nature 324, 446–449.

Benz, W., Asphaug, E., 1999. Catastrophic disruptions revisited. Icarus 142, 5–20.

Bottke Jr, W.F., Durda, D.D., Nesvorný, D., Jedicke, R., Morbidelli, A., Vokrouhlický, D., Levison, H.F., 2005. Linking the collisional history of the main asteroid belt to its dynamical excitation and depletion. Icarus 179, 63–94.

Citron, R.I., Genda, H., Ida, S., 2015. Formation of Phobos and Deimos via a giant impact. Icarus 252, 334–338.

Canup, R.M., Asphaug, E., 2001. Origin of the moon in a giant impact near the end of the earth's formation. Nature 412, 708–712.

Chambers, J., Wetherill, G., 1998. Making the terrestrial planets: N-body integrations of planetary embryos in three dimensions. Icarus 136, 304–327.

Durda, D.D., Greenberg, R., Jedicke, R., 1998. Collisional models and scaling laws: A new interpretation of the shape of the main-belt asteroid size distribution. Icarus 135, 431–440.

Fukushige, T., Makino, J., Kawai, A., 2005. GRAPE-6A: A single-card GRAPE-6 for parallel PC–GRAPE cluster systems. Publ. Astron. Soc. Japan 57, 1009–1021.

Genda, H., Kokubo, E., Ida, S., 2012. Merging criteria for giant impacts of protoplanets. Astrophys. J. 744, 137–144.

Goldreich, P., Ward, W.R., 1973. The formation of planetesimals. Astrophys. J. 183, 1051–1062.

Greenberg, R., Wacker, J.F., Hartmann, W.K., Chapman, C.R., 1978. Planetesimals to planets: Numerical simulation of collisional evolution. Icarus 35, 1–26.

Hayashi, C., Nakazawa, K., Nakagawa, Y., 1985. Formation of the Solar System. In: Black, D.C., Matthews, M.S. (Eds.), Protostars and Planets II. Univ. of Arizona Press, Tucson, pp. 1100–1153.

Hernquist, L., Katz, N., 1989. TREESPH: A unification of SPH with hierarchical tree method. Astrophys. J. Suppl. Ser. 70, 419–446.

Holsapple, K., Giblin, I., Housen, K., Nakamura, A., Ryan, E., 2002. Asteroid impacts: Laboratory experiments and scaling laws, In: Binzel, R.P., Bottke, W.F., Cellino, A., Paolicchi, P. (Eds.), Asteroids III. Tucson: Univ. of Arizona, pp. 443–462.

Hosono, N., Saitoh, T.R., Makino, J., 2013. Density-independent smoothed particle hydrodynamics for a non-ideal equation of state. Publ. Astron. Soc. Jpn. 65, 108(1)–108(11).

Housen, K.R., Holsapple, K.A., 1990. On the fragmentation of asteroids and planetary satellites. Icarus 84, 226–253.

Housen, K.R., Holsapple, K.A., 1999. Scale effects in strength-dominated collisions of rocky asteroids. Icarus, 142, 21–33.

Ida, S., Lin, D.N.C., 2004. Toward a deterministic model of planetary formation. I. A desert in the mass and semimajor axis distributions of extrasolar planets. Astrophys. J. 604, 388–413.

Inaba, S., Wetherill, G. W., Ikoma, M., 2003. Formation of gas giant planets: Core accretion models with fragmentation and planetary envelope. Icarus 166, 46–62.

Jutzi, M., Michel, P., Benz, W., Richardson, D.C., 2010. Fragment properties at the catastrophic disruption threshold: The effect of the parent body's internal structure. Icarus 207, 54–65.

Jutzi, M. 2015. SPH calculations of asteroid disruptions: The role of pressure dependent failure model. Planetary and Space Science 107, 3–9.

Kataoka, A., Tanaka, H., Okuzumi, S., Wada, K., 2013. Static compression of porous dust aggregates. Astron. Astrophys. 554, 1–12.

Kenyon, S.J., Bromley, B.C., 2008. Variations on debris disks: Icy planet formation at 30–150 AU for 1–3 $M_{sun}$ main-sequence stars. Astrophys. J. Suppl. Ser. 179, 451–483.

Kenyon, S.J., Bromley, B.C., 2012. Coagulation calculations of icy planet formation at 15–150 AU: A correlation between the maximum radius and the slope of the size distribution for trans-neptunian objects. Astrophys. J. 143, 63–83.

Kitsionas, S., Whitworth, A.P., 2002. Smoothed Particle Hydrodynamics with particle splitting, applied to self-gravitating collapse. Mon. Not. R. Astron. Soc. 330, 129–136.

Kobayashi, H., Dauphas, N., 2013. Small planetesimals formed Mars. Icarus 225, 122–130.

Kobayashi, H., Lohne, T., 2014. Debris disc formation induced by planetary growth. Man. Not. R. Astrom. Soc. 442, 3266–3274.

Kobayashi, H., Ormel, C.W., Ida, S., 2012. Rapid formation of Saturn after Jupiter completion. Astrophys. J. 756, 70 (7pp.).

Kobayashi, H., Tanaka, H., 2010. Fragmentation model dependence of collision cascades. Icarus 206, 735–746.

Kobayashi, H., Tanaka, H., Krivov, A.V., Inaba, S., 2010. Planetary growth with collisional fragmentation and gas drag. Icarus 209, 836–847.

Kobayashi, H., Tanaka, H., Krivov, A.V., 2011. Planetary core formation with collisional fragmentation and atmosphere to form gas giant planets. Astrophys. J. 738, 35–45.

Kokubo, E., Genda, H., 2010. Formation of terrestrial planets from protoplanets under a realistic accretion condition. Astrophys. J. L. 714, 21–25.

Kokubo, E., Ida, S., 1996. On runaway growth of planetesimals. Icarus 123, 180–191.

Leinhardt, Z.M., Ogilvie, G.I., Latter, H.N., Kokubo, E., 2012. Tidal disruption of satellites and formation of narrow rings. Mon. Not. R. Astron. Soc. 424, 1419–1431.

Leinhardt, Z.M., Richardson, D.C., 2002. *N*-body simulations of planetsimal evolution: Effect of varying impactor mass ratio. Icarus 159, 306–303.

Leinhardt, Z.M., Stewart, S.T., 2009. Full numerical simulations of catastrophic small body collisions. Icarus 199, 542–559.

Leinhardt, Z.M., Stewart, S.T., 2012. Collisions between gravity-dominated bodies. I. outcome regimes and scaling laws. Astrophys. J. 745, 79–105.

Love, S.G., Ahrens, T.J., 1996. Catastrophic impacts on gravity dominated asteroids. Icarus 124, 141–155.

Mizutani, H., Takagi, Y., Kawakami, S., 1990. New scaling laws on impact fragmentation. Icarus 87, 307–326.

Melosh, H.J., Ryan, E.V., 1997. Asteroids: Shattered but not dispersed. Icarus 129, 562–564.

Monaghan, J.J., 1992. Smoothed Particle Hydrodynamics. Annu. Rev. Astron. Astrophys. 30, 543–574.

Nakamura, A.M., Hiraoka, K., Yamashita, Y., Machii, N., 2009. Collisional disruption experiments of porous targets. Planetary and Space Science 57, 111–118.

Price, D.J., Monaghan, J.J., 2007. An energy-conserving formalism for adaptive gravitational force softening in smoothed particle hydrodynamics and N-body codes. Mon. Not. R. Astron. Soc. 374, 1347–1358.

Safronov, V.S., 1969. Evolution of the Protoplanetary Cloud and Formation of the Earth and the Planets. Nauka, Moscow.

Sekine, Y., Genda, H., 2012. Giant impacts in the Saturnian system: A possible origin of diversity in the inner mid-sized satellites. Planet. Space Sci. 63, 133–138.

Springel, V., 2010. Smoothed particle hydrodynamics in astrophysics. Annu. Rev. Astron. Astrophys. 48, 391–430.

Tillotson, J.H., 1962. Metallic Equations of State for Hypervelocity Impact. General Atomic Rept. GA-3216, 1–142.

Wyatt, M.C., Smith, R., Greaves, J.S., Beichman, C.A., Bryden, G., Lisse, C.M., 2007. Transience of hot dust around Sun-like stars. Astrophys. J. 658, 569–583.

Wada, K., Tanaka, H., Suyama, T., Kimura, H., Yamamoto, T., 2007. Numerical simulation of dust aggregate collisions. I. compression and disruption of two-dimensional aggregates. Astrophys. J. 661, 320–333.

Wada, K., Tanaka, H., Suyama, T., Kimura, H., Yamamoto, T., 2008. Numerical simulation of dust aggregate collisions. II. compression and disruption of three-dimensional aggregates in head-on collisions. Astrophys. J. 677, 1296–1308.

Wada, K., Tanaka, H., Suyama, T., Kimura, H., Yamamoto, T., 2009. Collisional growth conditions for dust aggregates. Astrophys. J. 702, 1490–1501.

Weidenschilling, S.J., 1980. Dust to planetesimals: Settling and coagulation in the solar nebula. Icarus 44, 172–189.

Weidenschilling, S.J., 1984. Evolution of grains in a turbulent solar nebula. Icarus 60, 553–567.

Wetherill, G.W., 1980. Formation of the terrestrial planets. Annu. Rev. Astron. Astrophys. 18, 77–113.

Wetherill, G.W., Stewart, G.R., 1989. Accumulation of a swarm of small planetesimals. Icarus 77, 330–357.

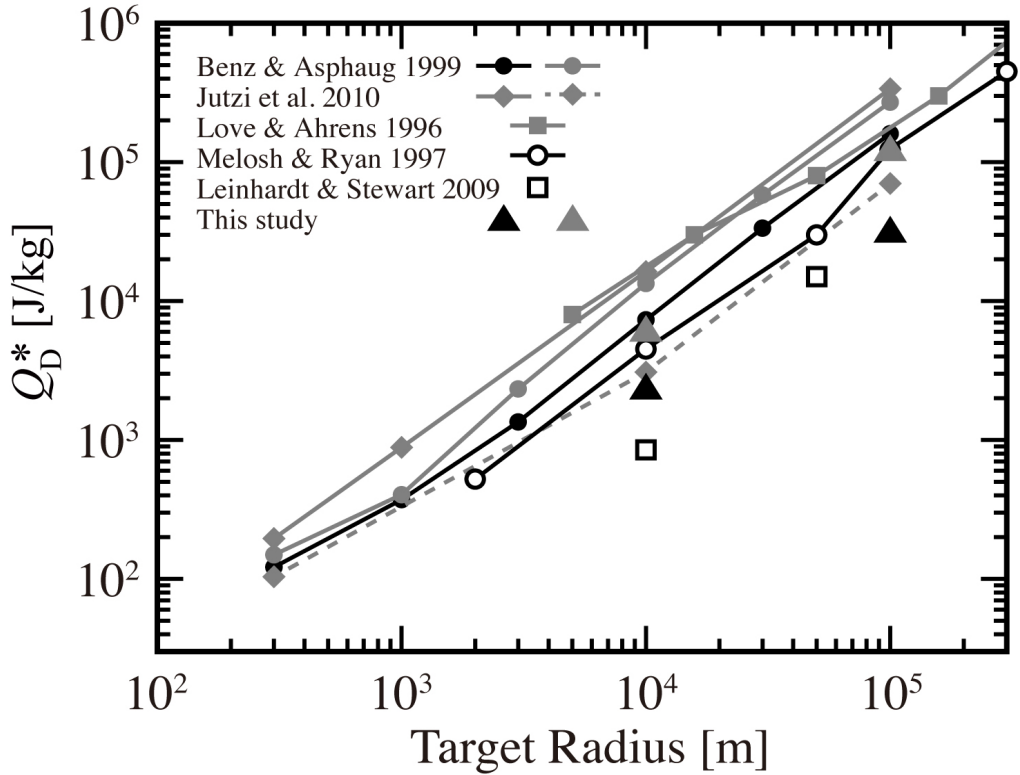

Figure 1. Critical specific impact energy for disruptive collision $Q_D^*$ for various target radii in the gravity regime calculated using several numerical methods. The results for collisions between basaltic objects (or granitic objects) are taken from the previous studies shown in the figure. Filled data points were obtained by SPH methods, and open data points by other numerical methods: the two-dimensional Lagrangian hydrocode (Melosh and Ryan, 1997) and the hybrid code of the Eulerian hydrocode and $N$-body code (Leinhardt and Stewart, 2009). Black and grey data points represent head-on and oblique (45°) collisions, respectively. Jutzi et al. (2010) considered target bodies with high (solid line) and low strengths (dashed line). Our results for the cases with the highest resolution ($5 \times 10^6$ SPH particles) are also shown in this figure (for details, see Section 3.3).

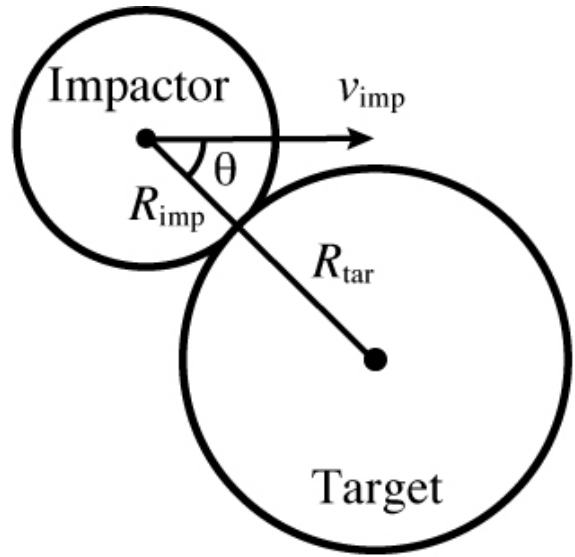

Figure 2. Geometry of a collision between a target and impactor with radii of $R_{tar}$ and $R_{imp}$ ($R_{tar} > R_{imp}$), respectively. The impact velocity and angle are $v_{imp}$ and $\theta$, respectively. A head-on collision corresponds to $\theta = 0°$.

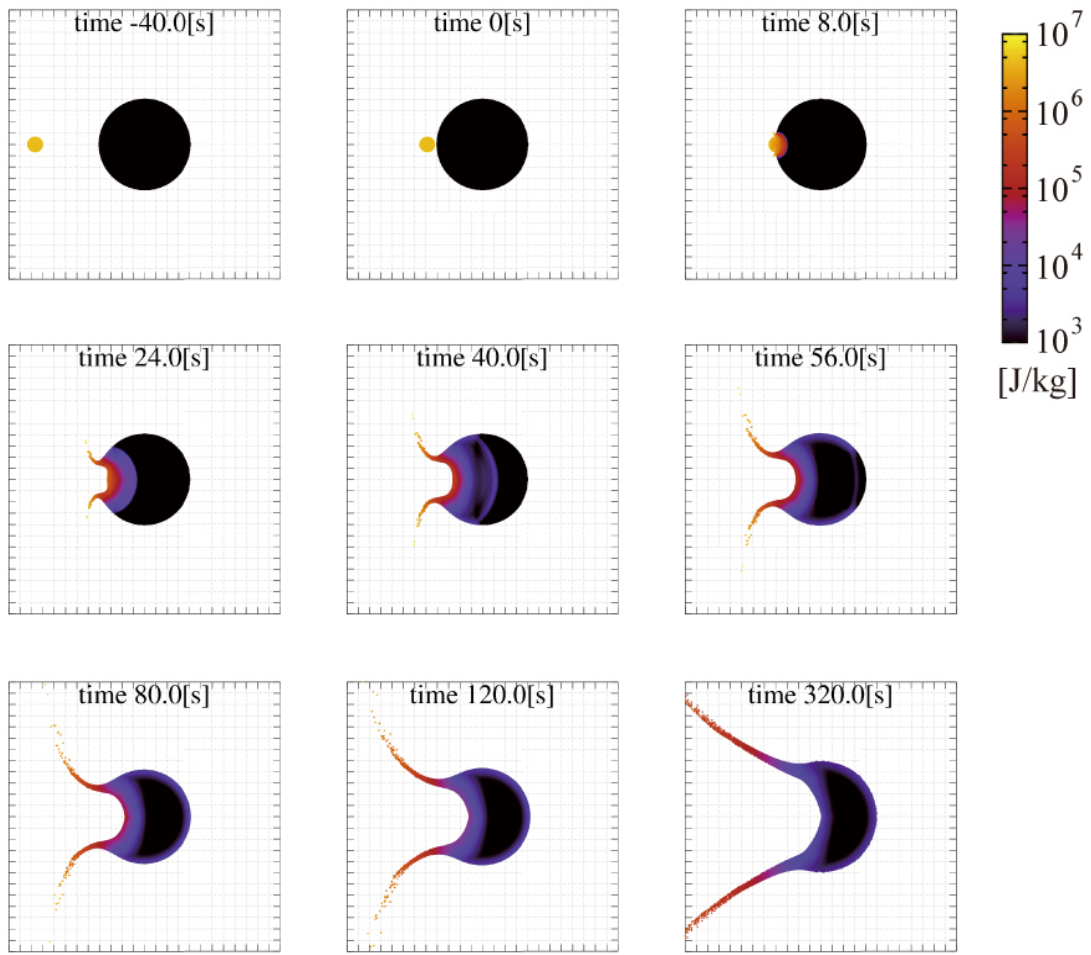

Figure 3. Snapshots of a head-on impact between a target with a 100 km radius and an impactor with a 15 km radius. The impact velocity is set to be 3 km/s. The color contour represents the specific kinetic energy.

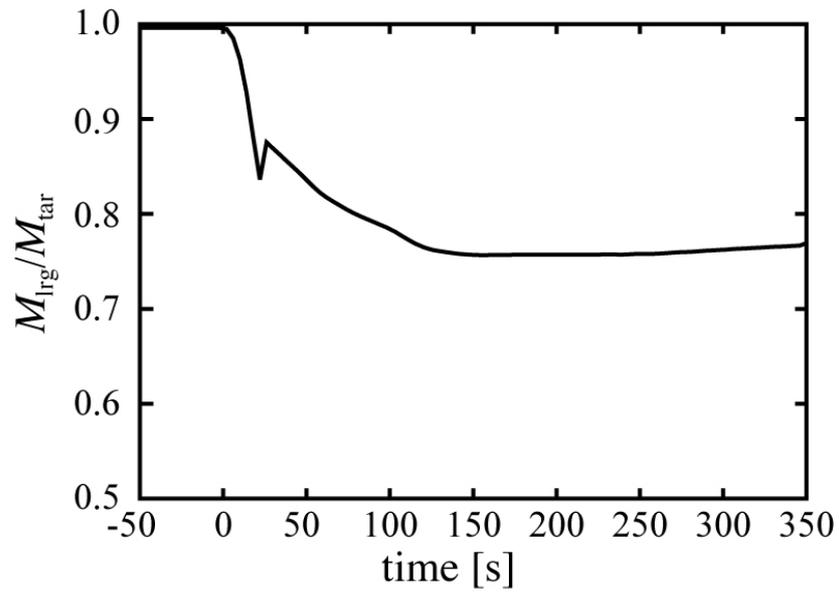

Figure 4. The evolution of the normalized mass of the largest body ($M_{lrg}/M_{tar}$). Contact between the two bodies occurs at $t = 0$ s. After the impact, $M_{lrg}/M_{tar}$ decreases steeply and almost converges after 150 s. The impact conditions are the same as those in Figure 3.

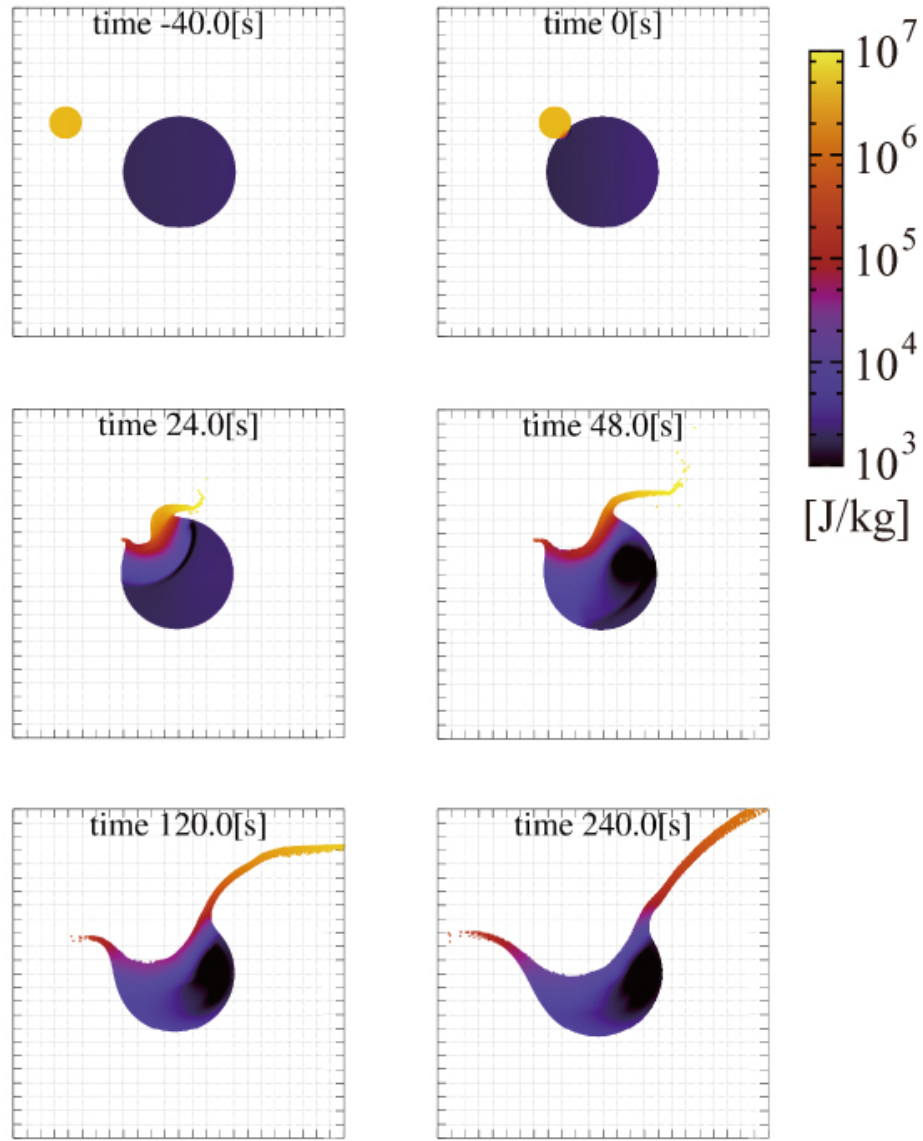

Figure 5. Snapshots of an oblique impact simulation. The target radius is 100 km, and the impactor radius is approximately 28 km. The impact velocity and the impact angle are 3 km/s and 45°, respectively. The color contour represents the specific kinetic energy.

- 23 -

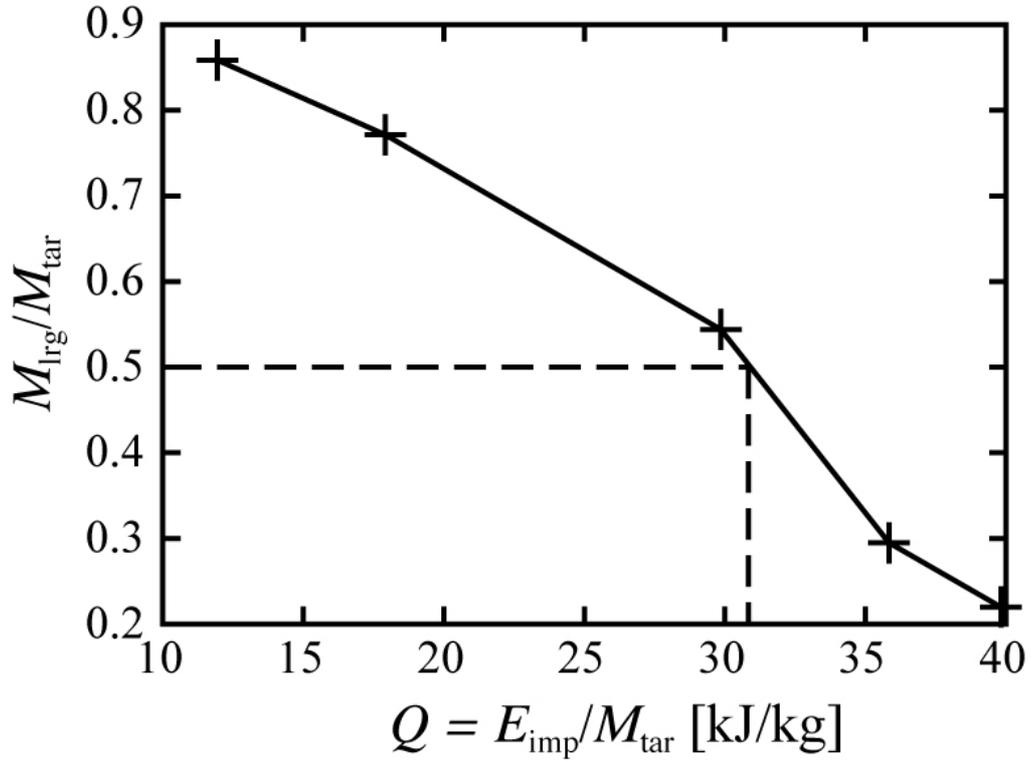

Figure 6. The mass of the largest body after collisions with different impact energies. The impact energy $E_{imp}$ is normalized by $M_{tar}$, and the mass of the largest body $M_{lrg}$ is normalized by the mass of the target $M_{tar}$. The five crosses represent the numerical results of our impact simulations. The impact parameters $R_{tar}$, $v_{imp}$, and $\theta$ are 100 km, 3 km/s, and 0°, respectively. The value of $Q_D^*$ can be calculated by linear interpolation of the data points and is estimated to be 31 kJ/s (dashed line).

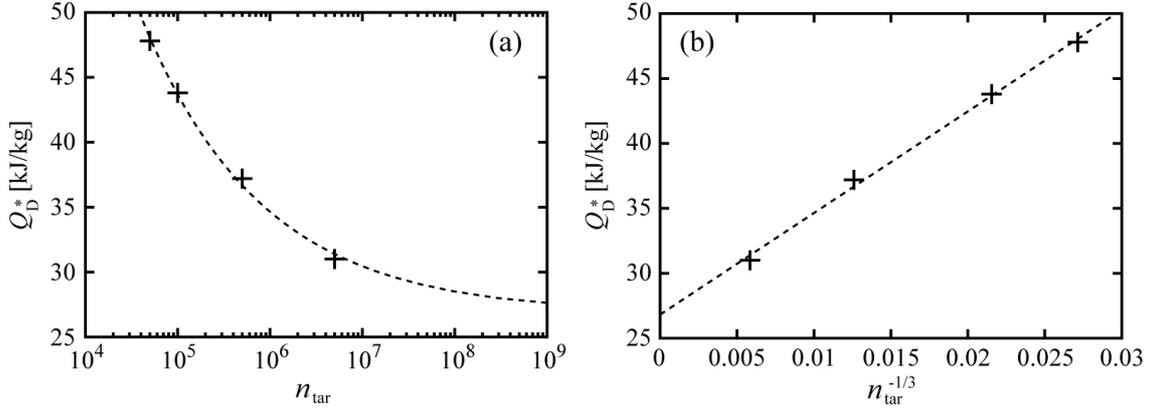

Figure 7. The dependence of $Q_D^*$ on the number of SPH particles used for the target with respect to (a) a log scale of the number of the particles $n_\mathrm{tar}$ and (b) the inverse of the number of particles per side $n_\mathrm{tar}^{-1/3}$. These $Q_D^*$ values are obtained by impact simulations under the same impact conditions with different numerical resolutions, where $R_\mathrm{tar}$, $v_\mathrm{imp}$, and $\theta$ are 100 km, 3 km/s, and 0°, respectively. The four crosses are our numerical results. The dashed curve is the best fit curve, which is given by Equation (1) and determined using four data points.

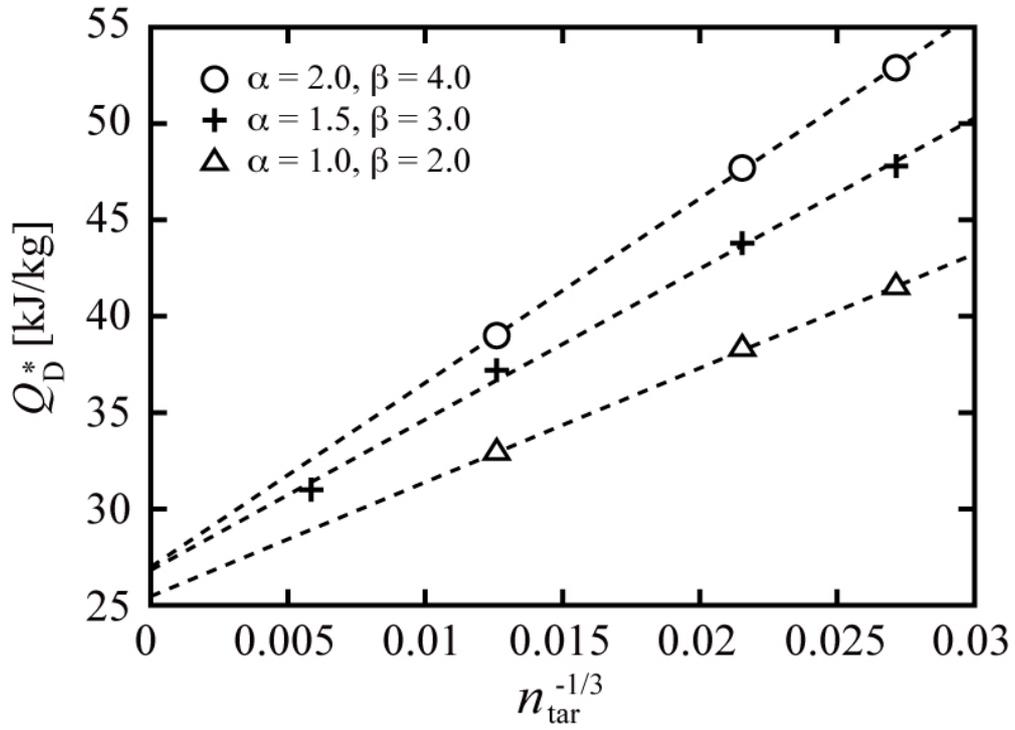

Figure 8. The dependence of $Q_D^*$ on the number of SPH particles and the artifical viscosity. Crosses are the same as those in Figure 7(b). Although the value of $Q_D^*$ depends on the value of the coefficient $\alpha$ and $\beta$ for a Von Neumann-Richtmyer-type viscosity, $Q_D^*$ seems to converge to the identical value.

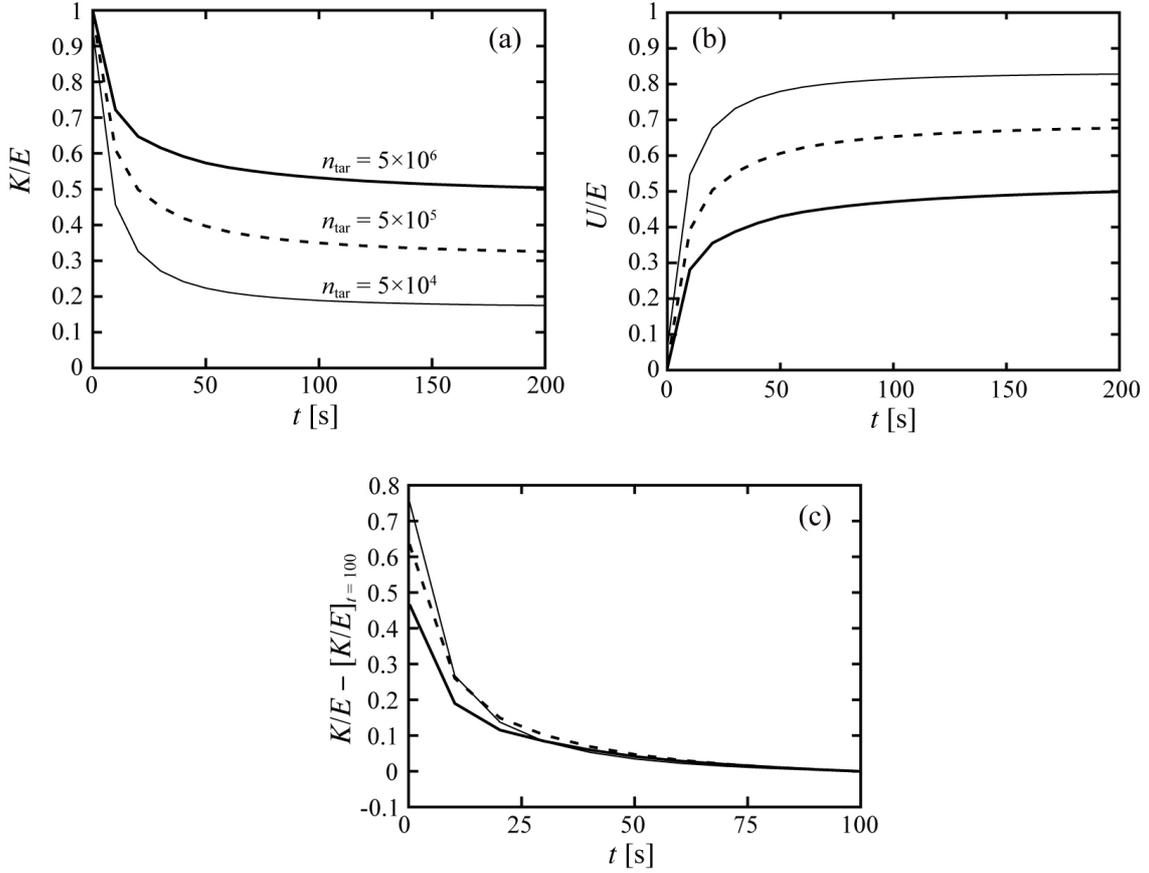

Figure 9. Evolution of (a) the total kinetic energy $K$ and (b) the total internal energy $U$ normalized by the total energy $E$, and (c) the difference between $K/E$ at each given time and that at $t = 100$ s. Solid, dashed, and thin solid curves correspond to impacts with $n_{\rm tar}$ = $5 \times 10^6$, $5 \times 10^5$, and $5 \times 10^4$, respectively. The impact conditions are the same as those in Figure 3 with the exception of the resolution. Contact of the impactor onto the surface occurs at $t = 0$ s. After the impact, part of the kinetic energy of the impactor is converted into the internal energy of the impactor and the target.

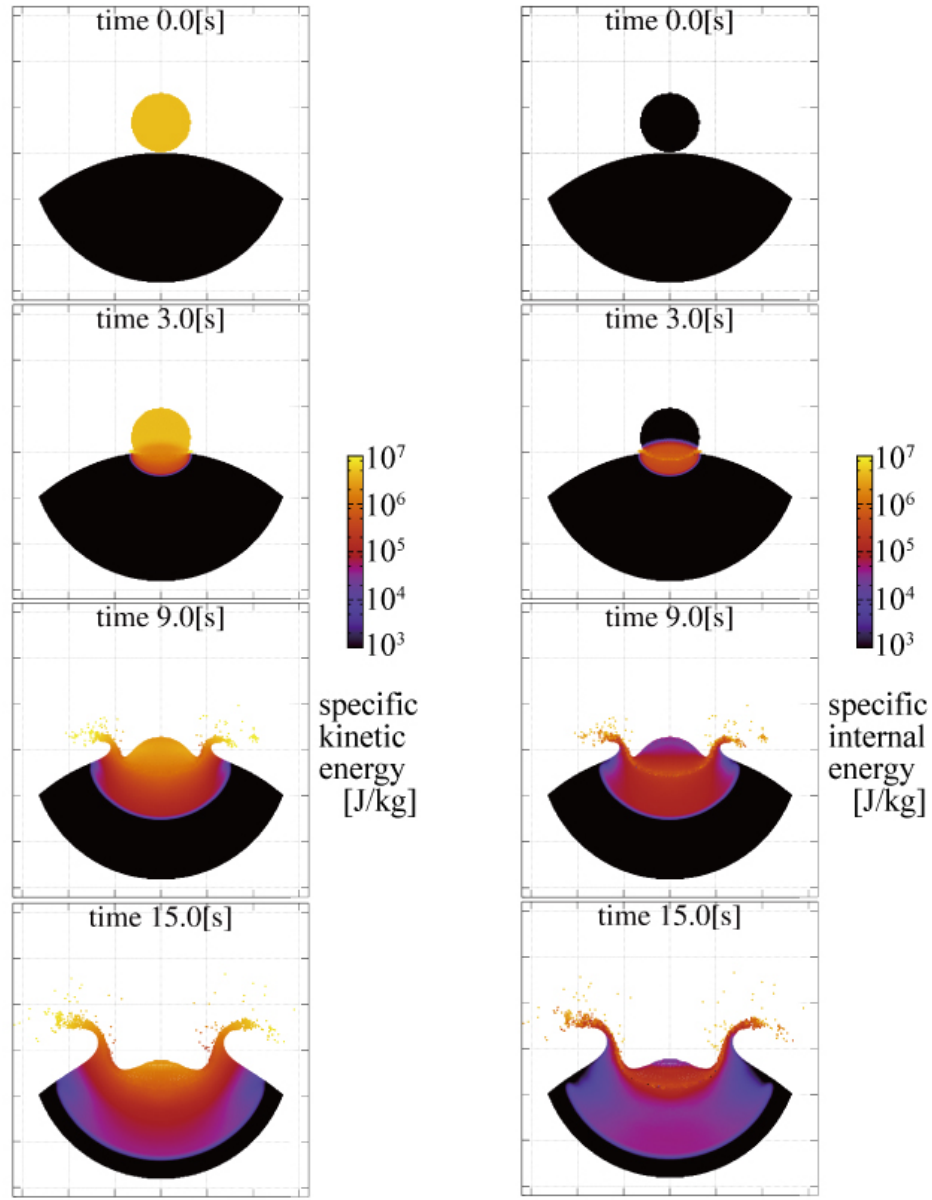

Figure 10. Snapshots of the cross section of a sphere-to-bowl impact simulation. The left four panels and the right four panels show the specific kinetic and specific internal energies, respectively. The topmost images are at $t = 0$ s. The impact velocity is 3 km/s, and the impactor radius is 15.8 km. The target bowl is part of a perfect sphere with a radius of 100 km. In total, $1.5 \times 10^7$ SPH particles are used, which correspond to $1.5 \times 10^8$ SPH particles for the sphere-equivalent numbers of SPH particles.

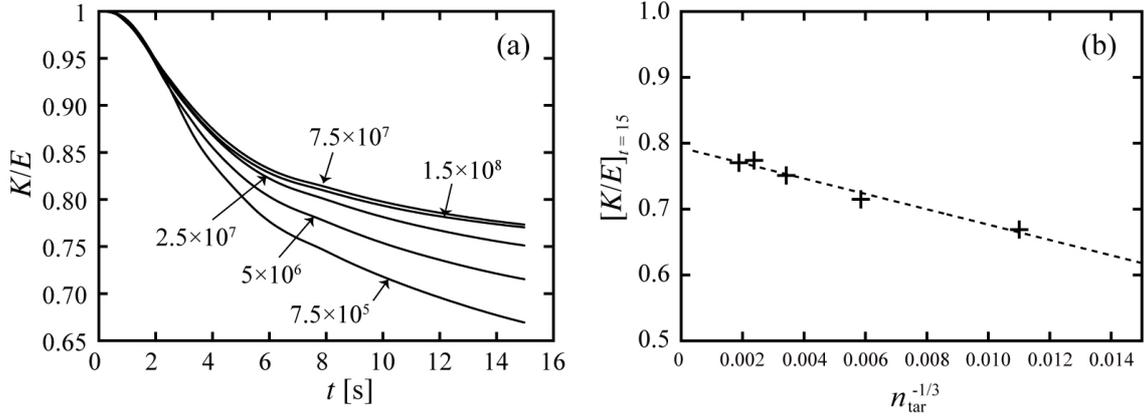

Figure 11. The temporal evolution of *K/E* in the sphere-to-bowl impact simulations with different resolutions (a), and the resolution dependence of *K/E* at 15 s (b). The numbers in (a) and $n_{tar}$ in (b) correspond to the sphere-equivalent numbers of SPH particles for the target. Impact conditions are the same as in Figure 9. The evolution of *K/E* in the simulations with $7.5 \times 10^7$ appears to be very similar to that with $1.5 \times 10^8$ particles. The dashed curve in (b) is the best fit curve, which is given by Equation (2) and determined using five data points.